\documentclass{aa}
\usepackage{graphics}

\begin{document}

   \thesaurus{03(03.13.6),08(08.05.03),11(11.13.1; 11.04.1)
              } 

   \title{The Tip of the Red Giant Branch and Distance of
          the Magellanic Clouds: results from the DENIS survey}

   \subtitle{}

   \author{Maria-Rosa L. Cioni\inst{1}
           \and
           Roeland P. van der Marel\inst{2}
	   \and
	   Cecile Loup\inst{3}
	   \and
	   Harm J. Habing\inst{1}
          }

   \offprints{mrcioni@strw.leidenuniv.nl}

   \institute{Sterrewacht Leiden, Postbus 9513,
              2300 RA Leiden, The Netherlands
         \and
	     Space Telescope Science Institute,
	     3700 San Martin Drive, Baltimore, MD 21218, USA
	 \and	
             Institute d'Astrophysique de Paris, CNRS, 
	     98 bis Bd. Arago, 75014 Paris, France
             }

   \date{Received 23 March 2000 / Accepted 5 May 2000}

   \titlerunning{The Tip of the Red Giant Branch and Distance of  
                 the Magellanic Clouds}
   \authorrunning{Maria-Rosa L. Cioni et al.}

   \maketitle

   \begin{abstract} We present a precise determination of the apparent
magnitude of the tip of the red giant branch (TRGB) in the $I$ ($0.8
\mu$m), $J$ ($1.25 \mu$m), and $K_S$ ($2.15 \mu$m) bands from the
luminosity function of a sample of data extracted from the DENIS
catalogue towards the Magellanic Clouds (Cioni et
al.~\cite{cio}). From the $J$ and $K_S$ magnitudes we derive
bolometric magnitudes $m_{bol}$. We present a new algorithm for the
determination of the TRGB magnitude, which we describe in detail and
test extensively using Monte-Carlo simulations. We note that any
method that searches for a peak in the first derivative (used by most
authors) or the second derivative (used by us) of the observed
luminosity function does not yield an unbiased estimate for the actual
magnitude of the TRGB discontinuity. We stress the importance of
correcting for this bias, which is not generally done. We combine the
results of our algorithm with theoretical predictions to derive the
distance modulus of the Magellanic Clouds. We obtain $m-M = 18.55 \pm
0.04$ (formal) $\pm 0.08$ (systematic) for the Large Magellanic Cloud
(LMC), and $m-M = 18.99 \pm 0.03$ (formal) $\pm 0.08$ (systematic) for
the Small Magellanic Cloud (SMC). These are among the most accurate
determinations of these quantities currently available, which is a
direct consequence of the large size of our sample and the
insensitivity of near infrared observations to dust extinction.
   \end{abstract}

\section{Introduction}
In the evolution of stars the position of the tip of the red giant branch 
(TRGB) marks the starting
point of helium burning in the core.  It is one of the strongest
characteristics of the life of stars seen in theoretical models,
together with the main sequence turn--off point, the red giant and the
asymptotic giant clump.  It has been used successfully for several
decades (Sandage \cite{san}) to estimate the distance of resolved
galaxies (e.g., Lee, Freedman \& Madore \cite{lfm}). The TRGB
magnitude depends only very weakly on age and metallicity, and yields
comparable precision as classical distance indicators such as Cepheids
and RR--Lyra variables.

Cioni et al.~(\cite{cio}) prepared the DENIS Catalogue towards the
Magellanic Clouds (DCMC), as part of the Deep Near Infrared Southern
Sky Survey performed with the 1m ESO telescope (Epchtein et
al.~\cite{epal}). The catalogue contains about $1\,300\,000$ and
$300\,000$ sources toward the LMC and the SMC, respectively; $70$\% of
them are real members of the Clouds and consist mainly of red giant
branch (RGB) stars and asymptotic giant branch (AGB) stars, and $30$\%
are galactic foreground objects.  This is a very large and homogeneous
statistical sample that allows a highly accurate determination of the
TRGB magnitude at the corresponding wavelengths. Among other things,
this yields an important new determination of the distance modulus of
the LMC.  This distance modulus is one of the main stepping stones in
the cosmological distance ladder, yet has remained somewhat uncertain
and controversial (e.g., Mould et al.~\cite{mould}).

Section~2 describes how the data were selected from the DCMC catalogue
to avoid crowding effects, and how we have calculated bolometric
corrections. Section~3 discusses the luminosity function (LF) and the
subtraction of the foreground component.  Section~4 discusses the TRGB
determination and gives comparisons with previous measurements.
Section~5 discusses the implications for the distances to the
Magellanic Clouds.  Concluding remarks are given in Section~6. The
Appendix provides a detailed description of the new method that we
have used to quantify the TRGB magnitude, as well a discussion of the
formal and systematic errors in the analysis.

\section{The Sample}

\subsection{The Data}                      
The DCMC covers a surface area of $19.87\times 16$ square degrees
centered on
$(\alpha,\delta)=(5^h27^m20^s$,$-69\degr00\arcmin00\arcsec)$ toward
the LMC and $14.7\times 10$ square degrees centered on
$(\alpha,\delta)=(1^h02^m40^s$,$-73\degr00\arcmin00\arcsec)$ toward
the SMC (J2000 coordinates).  We extracted all the sources detected
simultaneously in the three DENIS photometric wave bands: $I$ ($0.8
\mu$m), $J$ ($1.25 \mu$m) and $K_S$ ($2.15 \mu$m). We excluded sources
that were detected in all three wave bands but at
different times (this can happen because DENIS strips 
overlap). The selection of sources that are present in all three wave
bands strongly reduces possible crowding effects that affect mostly
the $I$ band.  We removed sources affected, even slightly, by image
defects (null image flag) and sources with bright neighbours or bad
pixels, sources that were originally blended, or sources with at least
one saturated pixel (null extraction flag). This increases the level
of confidence on the resulting sample. The main final sample for the
present analysis contains $33\,117$ sources toward the SMC and
$118\,234$ sources toward the LMC. This constitutes about $10$\% of
all the sources listed for each Cloud in the DCMC.

To estimate the contribution of the foreground component we also
considered the data in offset fields outside the spatial limits of the
DCMC\footnote{These data are not part of the DCMC catalogue but are
available on request from the first author.}, covering the same range
in right ascension and from a maximum of $\delta = -57^{\circ}$ to a
minimum of $\delta = -87^{\circ}$ (the full declination range of a
DENIS strip). These data were reduced together and the same selection
criteria, on the basis of the detection wave bands and the flags, were
applied as to the data constituting the DCMC. The total sample (DCMC
plus extension in declination) contains $92\,162$ and $184\,129$
sources in the RA ranges for the SMC and the LMC, respectively.

\begin{figure}
\resizebox{\hsize}{!}{\includegraphics{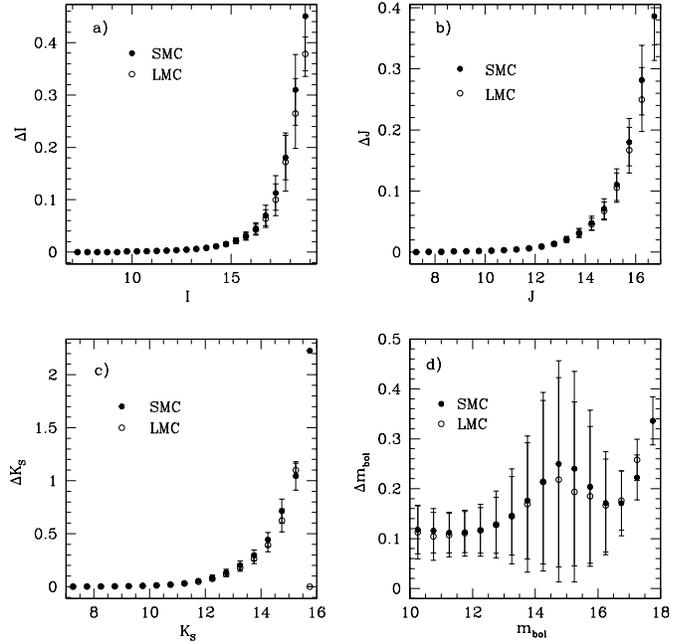}}
\caption{Distribution of the photometric errors. (a) $I$ band, (b) $J$
band, (c) $K_S$ band, (d) $m_{bol}$. Black dots are for sources toward
the LMC and empty dots are for sources toward the SMC. Error bars show
the dispersion in the photometric errors in $0.5$ mag bins.}
\label{errs}
\end{figure}

The distribution of the formal photometric errors in each wave band is
shown in Fig.~\ref{errs}. At the brighter magnitudes (those of
interest for the TRGB determination), the random errors in the sample
are not dominated by the formal photometric errors, but by
random errors in the photometric zero--points for the individual strips. 
The dispersions ($1 \sigma$) of these
zero-point variations are $0.07$ mag in the $I$ band, $0.13$ mag in
the $J$ band and $0.16$ mag in the $K_S$ band. Note that the formal
error with which the TRGB magnitude can be determined is not limited
to the size of these zero-point variations, but instead can be quite
small (the formal error is proportional to $1/\sqrt{N}$, where $N$ is
the number of stars in the sample).

The $I$, $J$ and $K_S$ magnitudes in the present paper are all in the
photometric system associated with the DENIS passbands. These
magnitudes are not identical to the classical Cousins $I$ and CTIO $J$
and $K$ magnitudes, although they are close (differences are $\leq
0.1$ magnitudes). The final transformation equations for the passbands
will not be available until the survey is completed, but a preliminary
analysis is presented by Fouqu\'e et al.~(\cite{fou}). Note that our
determinations of the distance moduli for the LMC and the SMC
(Section~5) are based on bolometric magnitudes derived from the data,
which are fully corrected for the specifics of the DENIS passbands.

\subsection{Bolometric correction}
We have calculated the apparent bolometric magnitude ($m_{bol}$) for
all the sources selected according to the criteria described in
Section~2.1, and with $(J-K_S) \ge 0.4$. We have chosen to use only
the $J$ and $K_S$ bands to derive $m_{bol}$ (see below). Sources with
$(J-K_S) < 0.4$ do not influence the position of the TRGB (see
Fig.~\ref{cm} below), and have too low a percentage of flux in the
near-infrared (NIR) to give a reliable measure of $m_{bol}$ with these
criteria.  We used two different bolometric corrections, depending on
the $(J-K_S)$ colour. For sources with $(J-K_S) < 1.25$, we simply use
a blackbody fit on the $(J-K_S)$ colour; such sources are mostly RGB
or early AGB (E--AGB) stars in our sample.  Sources with larger values
of $(J-K_S)$ are mostly thermally pulsing AGB (TP--AGB) stars, some of
which are losing mass and are surrounded by a circumstellar
envelope. For them we used the results of individual modelling of
galactic carbon (C) stars by Groenewegen et al.~(\cite{groen}),
combined with a series of models of increasing dust opacity where the
central star has a spectral type $M5$ and the dust grains are composed
of silicates (Groenewegen, private communication).

In both cases, blackbody fit and spectral models, our method to infer
$m_{bol}$ is different from what is usually performed in the
literature. We do not make any attempt to transform a magnitude,
i.e.~an integrated flux over the passband, into a flux density at a
reference wavelength, in order to suppress one step which already
makes an assumption on the spectral distribution of the source. We
only use the integrated flux measured over the $J$ and $K_S$ DENIS
passbands.  Theoretical spectral distributions, i.e.~blackbodies with
temperatures ranging from $10,000$ to $300$ K and the models from
Groenewegen and collaborators, were multiplied with the DENIS
passbands (which includes a mean atmosphere at la Silla observatory)
to derive the percentage of the total flux which is measured in each
DENIS passband as a function of the DENIS colours. Then, for each
selected DCMC source, $m_{bol}$ is calculated by interpolating in the
theoretical grids the percentage of flux measured in the $J$ and $K_S$
bands from the observed $(J-K_S)$ colour. We have used here the same
zero point as in Montegriffo et al.~(\cite{monte}). More details are
provided in Loup et al.~(\cite{lou}).

We have compared our results with the bolometric corrections $BC_K$
inferred by Montegriffo et al. As can be seen in their Fig.~3, their
bolometric correction is valid only for sources with $0.2 < (J-K_S) <
0.7$, with a typical spread around the fit of $0.1$ magnitude.  For
sources with $0.4 < (J-K_S) < 0.5$ our blackbody fit agrees with their
bolometric corrections to within the errors. On the other hand, for
some sources with $ (J-K_S) > 0.5$, they underestimate $m_{bol}$ by
$0.5$ to $2$ magnitudes compared to our calculations. This is not
surprising and can be inferred already from their Fig.~3; it does not
indicate a shortcoming in our approach. We also compared our results
with what one obtains by making blackbody fits using both the ($I-J$)
and ($J-K_S$) colours. For sources with $0.4 < (J-K_S) < 1.25$ it does
not produce any systematic effect; there is merely a spread of
typically $0.1$ magnitude between both calculations, consistent with
the formal errors. Inclusion of the $I$ band would produce a
systematic effect for bluer sources than those selected here, but
those are not relevant for the TRGB determination. We therefore
decided to use only the $J$ and $K_S$ band data in our calculations of
$m_{\rm bol}$, to minimize the effects of the interstellar reddening
which are much more pronounced in the $I$ band than in $J$ and $K_S$.

There are both random and systematic errors in our estimates of
$m_{\rm bol}$. The random errors come from two sources, namely from
the observational uncertainties in the observed $J$ and $K_S$ band
magnitudes, and from the corresponding uncertainties in the $(J-K_S)$
color. We have calculated the resulting random errors in the $m_{\rm
bol}$ estimates through propagation of these errors. There are also
two sources of systematic error in the $m_{\rm bol}$ estimates. The
first one derives from uncertainties in the dust extinction
correction. Our treatment of dust extinction is discussed in
Section~2.3; Appendix A.3.4 discusses how the uncertainties in this
correction introduce a small systematic error on the TRGB magnitude
determination. 
The second source of systematic
error comes from the difference between the real spectral
energy distribution of the star and the one we assume
to estimate $m_{bol}$. For blackbody fits, we did not make
any attempt to estimate this error because we lack information
for that purpose (we would need spectra and/or UBVRIJHKL
photometry on a sample of stars). For the AGB star models
from Groenewegen and collaborators, we can estimate part
of this error. The $(J-K_S)$ colour does not provide enough
information to fully constrain the set of model parameters,
i.e. $(J-K_S)$ does not give a unique solution, especially
when the chemical type of the star is unknown. With the models
available in this work, we have estimated this systematic
model error to be 5\% on the interpolated percentage of flux.
This is of course a lower limit as there can be some objects
whose spectral energy distribution differs from all
the ones produced in the models. On the
other hand, for most stars near the TRGB the blackbody fit is the
relevant model, and for these the systematic errors could be smaller.
In the end we have included in our final error budget a systematic
error of $\pm 0.05$ mag in our $m_{\rm bol}$ estimates due to
uncertainties in the underlying spectral model, but it should be noted
that this estimate is not very rigorous.

In our analysis of the TRGB magnitude we have propagated the 
random and systematic errors on $m_{bol}$ separately.
However, for illustrative purposes we
show in Fig.~\ref{errs}d the combined error. The surprising shape of
the error on $m{bol}$ as a function of $m_{\rm bol}$ 
should not be taken as real. It is an
artifact coming from the fact that a systematic model error was
included in the figure only for TP-AGB stars. The great majority of the
brightest stars are TP-AGB stars for which we use AGB models.
Going towards fainter stars, the $(J-K_S)$ colour decreases
and we mostly use blackbody fits, for which we have not included a 
systematic model error in the figure. The error on $m_{bol}$
thus seems to decrease around the TRGB.

\subsection{Dust Extinction}
The contribution of the internal reddening for the Magellanic Clouds is on
average only $E(B-V)=0.06$ while the foreground reddening can be very
high in the outskirts of the Clouds. We have not attempted to correct
our sample for extinction on a star by star basis. Instead we correct
all data for one overall extinction. We adopt $E(B-V)=0.15\pm0.05$ as
the average of known measurements (Westerlund \cite{west}) for both
Clouds. Adopting the extinction law by Glass (\cite{glass}) for the
DENIS pass bands [$A_V$\,:\,$A_I$\,:\,$A_J$\,:\,$A_{K_S}$ =
$1$\,:\,$0.592$\,:\,$0.256$\,:\,$0.089$] and $R_v=3.1$ we obtain
$A_I=0.27$, $A_J=0.11$ and $A_{K_S}=0.04$. Our approach to correct for
dust extinction is a simple approximation to what is in reality a very
complicated issue (e.g., Zaritsky~\cite{zar}). We discuss the effect
of uncertainties in the dust extinction on our results in Sections~4.4
and~5. While this is an important issue in the $I$ band, the bolometric
magnitudes that we use to determine the distance modulus are impacted
only at a very low level.
 
\section{The Luminosity Function}
The luminosity function (LF) of a stellar population is a powerful tool to
probe evolutionary events and their time scales. Major characteristics
of a stellar population are associated to bumps, discontinuities and
slope variations in the differential star counts as a function of
magnitude. However, for a proper interpretation of observed luminosity
functions several important issues should be taken into account. These
include the completeness of the sample of data, the foreground
contamination with respect to the analyzed population, the photometric
accuracy and the size of the sampling bins. The total number of
objects involved plays an important role to make the statistics
significant.

In most previous studies of the luminosity functions of stellar
populations in clusters or galaxies, in either the optical or the NIR,
limited statistics have been the main problem. The DENIS (Cioni et
al.~\cite{cio}) and 2MASS (Nikolaev \& Weinberg~\cite{Nik}) samples
provide the first truly large statistical sample in the NIR of the
Magellanic Cloud system. This wavelength domain is the most suitable
to study late evolutionary stages such as the RGB and the AGB. In the
present paper we restrict the discussion of the luminosity function
mostly to the TRGB; a more general discussion is given elsewhere
(Cioni, Messineo \& Habing 2000).

\subsection{The contribution of the Galaxy}
\begin{figure}
\resizebox{\hsize}{!}{\includegraphics{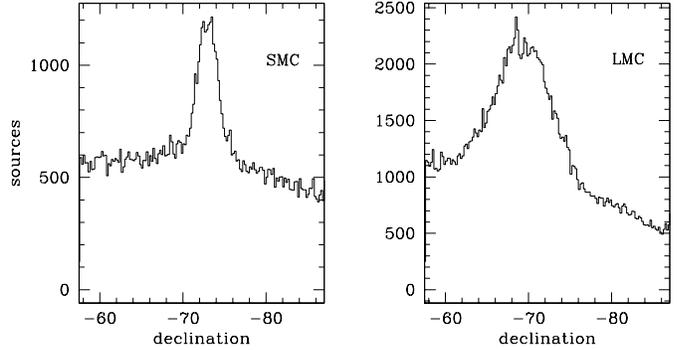}}
\caption{Distribution of the sources in the sample versus declination
using bins of $0.1$ degrees, for the SMC (left) and the LMC (right).}
\label{dec}
\end{figure}

For the removal of foreground contamination we considered two offset
fields around each cloud. The range of right ascension (RA) is the same 
for both the cloud and the offset fields; it is the same of the DCMC 
catalogue (Section 2.1). 
For the LMC the north field has $-58^{\circ}
> \delta >-60^{\circ}$ and the south field has $-80^{\circ} > \delta >
-86^{\circ}$; for the SMC the north field has $-60^{\circ} > \delta >
-66^{\circ}$ and the south field has $-80^{\circ} > \delta >
-86^{\circ}$. The LMC region itself was limited to the declination
range $-62^{\circ} > \delta > -76^{\circ}$, and the SMC region to
$-69^{\circ} > \delta > -77^{\circ}$. Fig.~\ref{dec} shows the
distribution versus declination of the sources in the sample, using
bins of $0.1$ degrees. The foreground contribution clearly decreases
toward more negative declinations, due to the difference in Galactic
latitude. The difference in number between the foreground contribution
around the LMC and around the SMC is consistent with the fact that
the LMC is observed closer to the galactic plane than the SMC is.  The
structure of the LMC is clearly wider than the one of the SMC and this
may contribute to create the strong declination trend around the LMC.

\begin{figure*}
\resizebox{0.8\hsize}{!}{\includegraphics{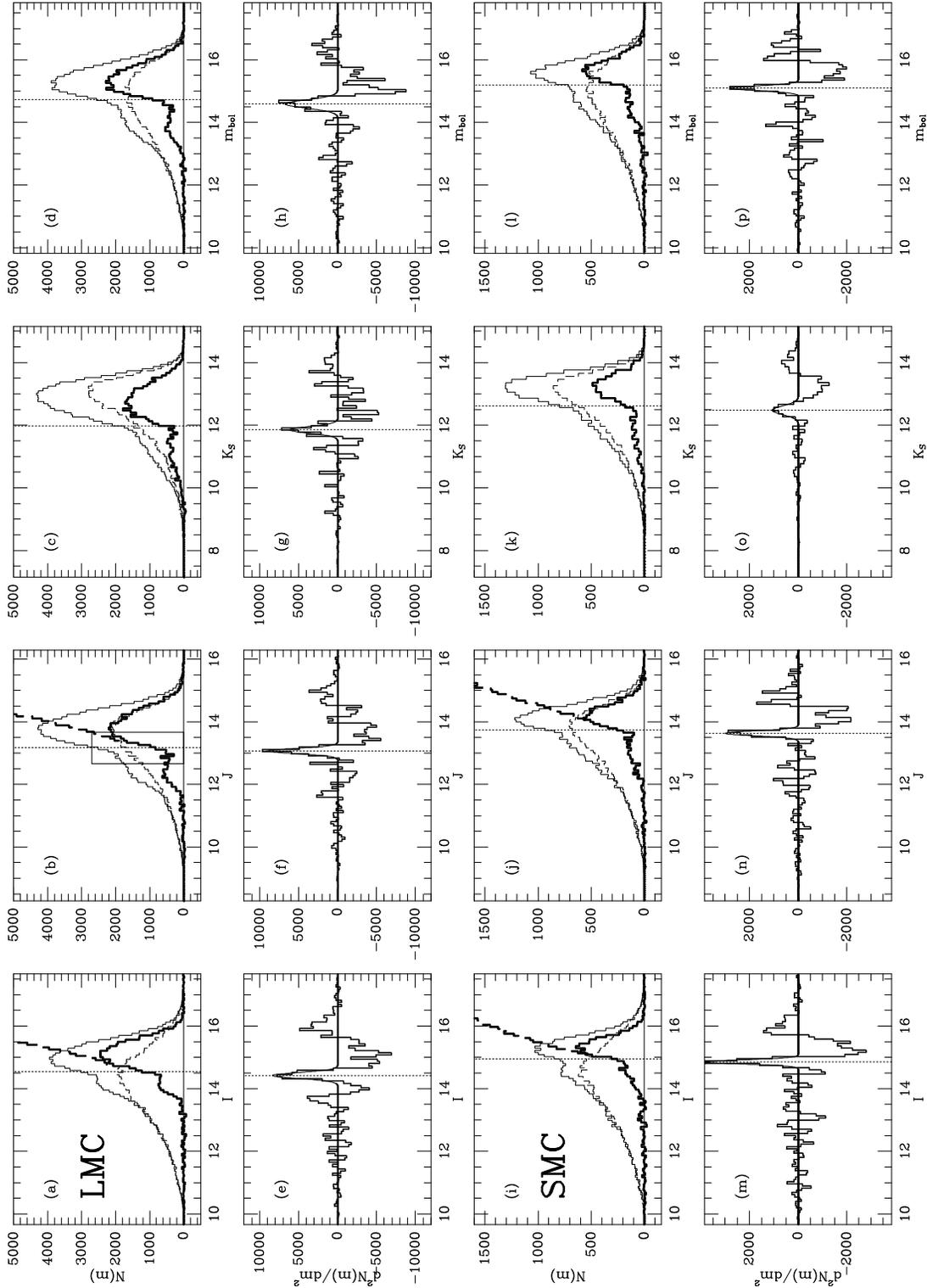}}
\caption{Stellar magnitude distributions, $N(m)$, and second
derivative after the application of a Savitzky-Golay filter, ${\rm
d}^2 N(m) / {\rm d}m^2$, for the LMC (a--h) and the SMC (i--p).
Panels (a--d) and (i--l) show the distributions for the main field
(thin solid curve), for the scaled offset field (thin dashed curve),
and for the foreground--subtracted main field (heavy solid curve).
For the $I$ and $J$ bands we also show the distribution for the
foreground--subtracted main field for the larger sample of all stars
detected in $I$ and $J$ (irrespective of $K_S$; heavy long--dashed
curves). The final estimate of the TRGB discontinuity is indicated
(vertical dotted line). The unit along the ordinate is the number of
stars per $0.07$ mag bin. Panels (e--h) and (m--p) show the second
derivative for the foreground--subtracted main field (heavy solid
curve), the best Gaussian fit to the peak (thin solid curve), and the
position of the peak (vertical dotted line). The solid rectangle in
(b) outlines the region shown in detail in Fig.~\ref{mod}.}
\label{tip}
\end{figure*}

For each field and photometric band we constructed a histogram of the
observed magnitudes (thin solid curves in the $N(m)$ panels of
Fig.~\ref{tip}).  For the two different offset fields at each right
ascension range the data were combined into one histogram.  This
offset--field histogram (thin dashed curves) was then scaled to fit
the corresponding LMC or SMC field histogram at bright magnitudes, for
which almost all the stars belong to the foreground. Subtraction
yields the foreground--subtracted magnitude distribution for each of
the Clouds (heavy solid curves).  For comparison we also extracted
from the catalogue an extended sample consisting of those stars
detected in the $I$ and $J$ bands (irrespective of whether or not they
were detected in $K_S$). This sample (heavy dashed curves) is complete
to fainter magnitudes than the main sample, and therefore illustrates
the completeness limit of the main sample.

\subsection{The shape}
The resulting statistics of the subtracted LF are impressive, despite
the restricted source selection.  We proceed with a description of the
major characteristics of the LF.  The maximum corresponds to giants
that lie on the upper part of the RGB. The decrease at fainter
magnitudes is due to the selections applied to the data and to the
decrease in sensitivity of the observations (Cioni et al.~\cite{cio}).
Features like the horizontal branch or the red clump are too faint to
be detected by DENIS. Towards brighter magnitudes we encounter a strong kink in
the profile, which we associate with the position of the TRGB
discontinuity. Brightward of the kink follows a bump of objects which
we discuss below. At very bright magnitudes the LF has a weak tail
which is composed of stars of luminosity type I and II 
(Frogel \& Blanco~\cite{fb}), but the LF at these bright magnitudes could 
be influenced by small residuals due to inaccurate foreground subtraction.

\begin{figure}
\resizebox{\hsize}{!}{\includegraphics{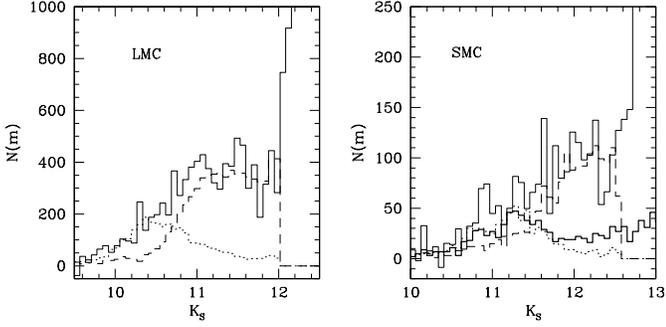}}
\caption{Differential count of the number of sources detected versus
magnitude in the area of the Magellanic Clouds after the subtraction
of the foreground contribution (thin solid line).
This enlarges part of Figs.~\ref{tip}c and
\ref{tip}k. The curves show the contributions of O--rich AGB stars
(dashed), C--rich TP--AGB stars (dotted), and spectroscopically
confirmed C--rich AGB stars (thick solid for the SMC only).}
\label{agb}
\end{figure}

To explain the bump brightward of the TRGB discontinuity we
cross--identified (Loup~\cite{loup}) the DCMC sources with the
sources in some of the Blanco fields in the LMC (Blanco \cite{bbmc}).
In the $(K_S, J-K_S)$ diagram there are two regions populated only by
oxygen rich AGB stars (O--rich) and by carbon rich AGB stars
(C--rich), respectively.  O--rich stars are concentrated around
$K_S=11.5$ and have a constant color $(J-K_S)=1.2$, and C--rich stars
are concentrated around $(J-K_S)=1.7$ and around $K_S=10.5$ (see
Fig.~\ref{cm}b).  These TP--AGB stars cause the bump visible in
the LF.  This bump should not be confused with the AGB bump caused by
E--AGB stars (Gallart \cite{gal}).  Fig.~\ref{agb} shows an
enlargement of Figs.~\ref{tip}c and \ref{tip}k (continuous line).  The
dashed line refers to O--rich AGB stars and the dotted line to C--rich
AGB stars selected in the $(K_S, J-K_S)$ diagram.  In the case of the
SMC we selected regions with slightly bluer color and fainter
magnitude to match the two groups of AGB stars in the $(K_S, J-K_S)$
diagram, cf.~Fig.~\ref{cm}d.  Fig.~\ref{agb} also plots the LF (thick
line) that results when we cross--identify our sample with the
spectroscopically confirmed carbon stars by Rebeirot et
al.~(\cite{reb}) in the SMC.  We found $1451$ sources out of $1707$
and we attribute the missing cross--identifications to the selection
criteria that we applied to the DCMC data to obtain the sample for the
present paper.  It is interesting to note that at higher luminosities
the distribution of the confirmed C--rich stars matches the
distribution of C--rich stars selected only on the basis of $K_S$ and
$(J-K_S)$.  At the fainter luminosities C--rich AGB stars cannot be
discriminated from O--rich AGB stars only on the basis of $(J-K_S)$
and $K_S$ because they overlap with the RGB, principally constituted
by O--rich stars.

\section{The tip of the RGB}

\subsection{Theory}
Theoretically stars climb the RGB with an expanding convective
envelope and an hydrogen burning shell, while increasing the
core--Helium content, the central temperature, the central density,
and the luminosity.  Low--mass stars ($0.8-1.0 M_\odot<M<2-2.3
M_\odot$) develop an electron--degenerate core, which causes an
explosive start (Helium--flash) of the core--Helium burning when the
core mass reaches $0.45 M_\odot$, almost independently from the
initial mass and composition of the star (Chiosi et al.~\cite{cbb});
intermediate mass stars ($2-2.3 M_\odot< M< 8-9 M_\odot$) are not
affected by degeneracy at this stage and initiate helium burning quietly, when
a suitable temperature and density are reached. The RGB transition
phase between the two behaviors occurs when the population is at least
$0.6$ Gyr old and lasts roughly for $0.2$ Gyr, determining an abrupt
event in the population life time (Sweigart et al.~\cite{sgr}).  The
Helium--flash is followed by a sudden decrease in the luminosity
because of the expansion of the central region of the star and because
of the extinction of the hydrogen--burning shell, the major nuclear
energy supply. The star reaches its maximum luminosity and radius (in
the RGB phase) at the TRGB, which also marks the end of the phase
itself (Iben \cite{iben}).  Low--mass stars with the same metallicity
accumulate along the RGB up to a TRGB luminosity of about $2500
L_\odot$ (Westerlund \cite{west}); the resulting RGB is quite
extended.  Stars with masses just above the transition mass (which
discriminates between low and intermediate masses) have a TRGB
luminosity as low as $200 L_\odot$ (Sweigart et al.~\cite{sgro},
\cite{sgr}) and the RGB is almost non--existent.  Both low and
intermediate mass stars that finish burning their Helium in the core
evolve on the AGB phase. They are in the so called E--AGB when Helium
is burning in a thick shell and in the so called TP--AGB when both the
Hydrogen and the Helium shells are active.  The luminosity increases
because of the increase in mass of the degenerate carbon core.  The
AGB evolution is characterized by a strong mass loss process that ends
the phase when the outer envelope is completely lost. The maximum AGB
luminosity defines the tip of the AGB (TAGB), with core mass
$M_{\rm core}=1.4 M_\odot$ and magnitude $M_{bol}=-7.1$ mag (Paczynski
\cite{pac}).

\begin{figure}
\resizebox{\hsize}{!}{\includegraphics{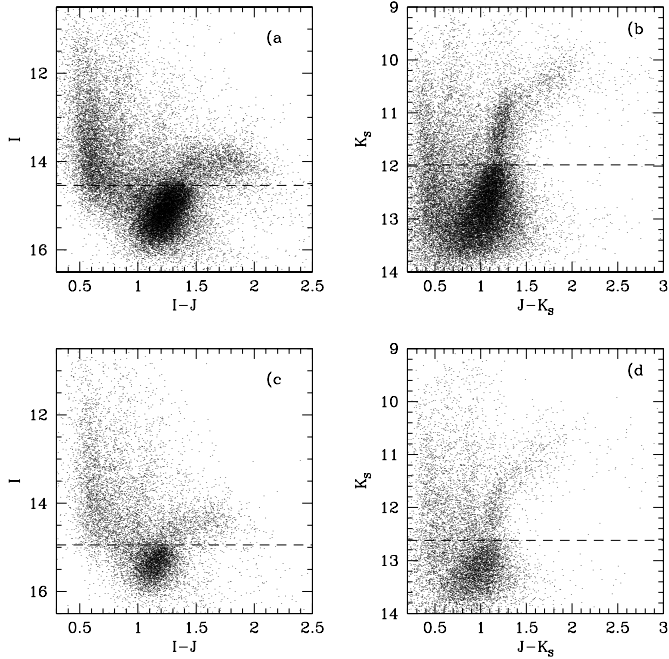}}
\caption{Color--magnitude diagrams of $(I, I-J)$ on the left and
$(K_S, J-K_S)$ on the right for sources detected toward the LMC with
$-67^{\circ} > \delta > -69^{\circ}$ (panels~a and~b) and toward the
SMC with $-72^{\circ} < \delta < -74^{\circ}$ (panels~c and~d). A
dashed horizontal line in each panel indicates the TRGB magnitude
derived in Section~4 (Table~\ref{value}).}
\label{cm}
\end{figure}

\subsection{Detections}
In the observed diagrams $(I, I-J)$ and $(K_S, J-K_S)$ the RGB is
clearly visible (Fig.~\ref{cm}). The beginning of the RGB phase is
below the detection limits and the spread at the fainter magnitudes is
due to the photometric errors. The TRGB is clearly defined at the
brightest point of this branch as an outstanding roughly horizontal
feature. Dashed horizontal lines in the figure indicate the values of
the TRGB discontinuity that we derive below for these data. The plume
of objects brighter than the TRGB is composed of AGB stars
experiencing the TP phase.  From these diagrams the foreground
contribution has not been subtracted but the contamination of these to
the RGB/AGB is negligible (Cioni et al.~\cite{chl}, \cite{cio}) if
only the very central region of each cloud is selected; Fig.~\ref{cm}
contains sources with $-67^{\circ} > \delta > -69^{\circ}$ toward the
LMC and $-72^{\circ} > \delta > -74^{\circ}$ toward the SMC.  Stars
populating the RGB up to the TRGB are low--mass stars older than $0.6$
Gyr. TP--AGB stars on the other hand, which lie above the TRGB, can be
either low--mass stars or intermediate mass--stars. For $M_{bol}<-6$
mag they all originate from main--sequence stars with $M<3 M_\odot$
(Westerlund \cite{west}), which corresponds to a minimum age of $0.2$
Gyr. TP--AGB stars that are low--mass stars should be older than $1$
Gyr (Vassiliadis and Wood \cite{vw}). Note that the thickness of the
RGB ($\sim 0.3$ mag) is larger than the photometric errors involved
($\sim 0.1$ mag) and this indicates a spread in either metallicity or
extinction within each cloud.

  \begin{table} \caption[]{Summary of TRGB magnitude determinations
      and errors. Column~(1): type of magnitude, i.e., either the
      photometric band or $m_{\rm bol}$. Listed magnitudes for 
      $I$, $J$ and $K_S$ are in the photometric system of the DENIS passbands 
      (Fouqu\'e et al.~\cite{fou}). Column~(2): Cloud name.
      Column~(3): observed magnitude of the TRGB (not corrected for
      extinction), determined using the algorithm described in
      Appendix~A. Column~(4): magnitude of the TRGB
      corrected for extinction under the assumption that $E(B-V) =
      0.15$. Column~(5): formal error in $m_{\rm TRGB}$ derived from
      Monte-Carlo simulations as described in Appendix~A.
      Column~(6): the amount by which the extinction-corrected $m_{\rm
      TRGB}$ would change if the assumed $E(B-V)$ were increased by
      $+0.05$ (a change of $-0.05$ yields the opposite change
      in $m_{\rm TRGB}$).}
         \label{value}
      \[
         \begin{array}{rrcccc}
            \hline
            \noalign{\smallskip}
     \mathrm{Type}&\mathrm{Cloud}&m_{\rm TRGB}&m_{\rm TRGB}&\Delta m_{\rm TRGB}&\delta_{\rm dust} \\
      & &\mathrm{(observed)}&\mathrm{(dereddened)}&\mathrm{(formal)}& \\
     (1)&(2)&(3)&(4)&(5)&(6) \\     
            \noalign{\smallskip}
            \hline
            \noalign{\smallskip}
     I           & {\rm LMC} &  14.54 & 14.27 & 0.03 & -0.09 \\
     I           & {\rm SMC} &  14.95 & 14.68 & 0.03 & -0.09 \\
     J           & {\rm LMC} &  13.17 & 13.06 & 0.02 & -0.04 \\
     J           & {\rm SMC} &  13.73 & 13.62 & 0.03 & -0.04 \\
     K_S         & {\rm LMC} &  11.98 & 11.94 & 0.04 & -0.02 \\
     K_S         & {\rm SMC} &  12.62 & 12.58 & 0.07 & -0.02 \\
     m_{\rm bol} & {\rm LMC} &  ---   & 14.73 & 0.04 & -0.03 \\
     m_{\rm bol} & {\rm SMC} &  ---   & 15.19 & 0.03 & -0.03 \\
            \noalign{\smallskip}
            \hline
         \end{array}
      \]
   \end{table}

\subsection{Method}
The algorithm that we have used for the determination of the position
of the magnitude $m_{\rm TRGB}$ of the TRGB is described in great
detail in Appendix~A. The TRGB discontinuity causes a peak in both the
first derivative $N'(m) \equiv {\rm d} N(m) / {\rm d} m$ and the
second derivative $N''(m) \equiv {\rm d}^2 N(m) / {\rm d} m^2$ of the
observed stellar magnitude distribution $N(m)$. Previous authors have
generally used $N'(m)$ to estimate $m_{\rm TRGB}$ (e.g., Madore \&
Freedman \cite{mado}). Based on extensive tests
and simulations we found that for our dataset $N''(m)$ provides a better handle
on $m_{\rm TRGB}$ (cf.~Appendix A.1). We therefore adopted the
following approach. First, we use a Savitzky-Golay filter (e.g., Press
et al.~\cite{press}) to estimate $N''(m)$. We then search for a peak
in $N''(m)$, and fit a Gaussian to it to obtain the quantities
$m_{2g}$ and $\sigma_{2g}$ that are the mean and dispersion of the
best-fitting Gaussian, respectively. The magnitude $m_{\rm TRGB}$ is
then estimated as $m_{2g} + \Delta m_{2g} (\sigma_{2g})$, where
$\Delta m_{2g} (\sigma_{2g})$ is a small correction (Fig.~\ref{mag}b)
derived from a phenomenological model described in Section~A.1. The
formal errors on the $m_{\rm TRGB}$ determinations are inferred from
extensive Monte-Carlo simulations, as described in Section~A.2. The
possible influence of systematic errors is discussed in
Section~A.3. There is no evidence for any possible systematic errors
due to possible incompleteness in the sample, or inaccuracies in the
foreground subtraction. Systematic errors due to uncertainties in the
phenomenological model on which the corrections $\Delta m_{2g}
(\sigma_{2g})$ are based can be up to $\pm 0.02$ magnitudes.
Extinction variations within the Clouds do not cause systematic errors
in either the estimate of $m_{\rm TRGB}$ or its formal error. However,
any error in the assumed {\it average} extinction for the sample does
obviously translate directly into an error in $m_{\rm TRGB}$.

Fig.~\ref{tip} summarizes the results of the analysis. The second and
fourth row of the panels show the estimates of $N''(m)$. The Gaussian
fit to the peak is overplotted, and its center $m_{2g}$ is indicated
by a vertical dotted line.  The corresponding estimate $m_{\rm TRGB}$
is indicated by a vertical dotted line in the panel for $N(m)$. Table
\ref{value} lists the results. It includes both the observed value for
$m_{\rm TRGB}$, as well as the value obtained after correction for
extinction with $E(B-V)=0.15$. Formal errors are listed as well, and
are typically $0.03$--$0.04$ magnitudes.  The last column of the Table
lists the amount by which the extinction-corrected $m_{\rm TRGB}$
would change if the assumed $E(B-V)$ were increased by $+0.05$ (a
shift of $-0.05$ in the assumed $E(B-V)$ would produce the opposite
shift in $m_{\rm TRGB}$).

When applying comparable methods to resolvable galaxies in the Local
Group (e.g., Soria et al \cite{soria}; Sakai et al.~\cite{sakai}) one
of the major sources of contamination on the TRGB determination is the
presence of a relative strong AGB population. The Magellanic Clouds
also have a strong AGB population, but in our case this does not
confuse the determination of $m_{\rm TRGB}$. This is due to the large
statistics available, and above all to the fact that TP--AGB stars are
definitely more luminous than the TRGB. E--AGB stars overlap with the
RGB stars but there is no reason to assume, according to models, that
they accumulate at the TRGB. Probably they distribute rather
constantly and due to the very short evolutionary time scale we do not
expect them to exceed more than $10$\% of the RGB population.

\subsection{Discussion}
The absolute magnitude of the TRGB generally depends on the metallicity
and the age of the stellar population 
and therefore need not to be the same for the LMC and the
SMC. Nonetheless, if we assume that such differences in TRGB absolute
magnitude are small or negligible, and if we assume that the
extinction towards the LMC and the SMC have been correctly estimated,
then one may subtract for each photometric band the inferred $m_{\rm
TRGB}$(LMC) from the inferred $m_{\rm TRGB}$(SMC) to obtain an
estimate of the difference $\Delta \equiv (m-M)_{\rm SMC} - (m-M)_{\rm
LMC}$ between the distance moduli of the SMC and the LMC. 
This yields the following results: $0.41 \pm 0.04$ ($I$ band), $0.56
\pm 0.04$ ($J$ band), $0.64 \pm 0.08$ ($K_S$ band) and $0.46 \pm 0.05$
($m_{\rm bol}$). The dispersion among these four numbers is $0.09$,
which is somewhat larger than the formal errors. Averaging the four
determinations yields $\Delta = 0.52 \pm 0.04$, where the error is
the formal error in the mean. This is not inconsistent with
determinations found in the literature, which generally fall in the
range $\Delta = 0.4$---$0.5$ (Westerlund \cite{west}).

Upon taking a closer look at the values of $\Delta$ for the different
bands one sees that the values in $J$ and $K_S$ exceed those in $I$ by
$0.15$ mag or more. It is quite possible that this is due to differences in
the metallicity and age of the LMC and the SMC, which affect the TRGB
absolute magnitude $M_{\rm TRGB}$ differently in different bands. In
the $I$ band $M_{\rm TRGB}$ is reasonably insensitive to metallicity
and age. Lee et al.~(\cite{lfm}) showed that $M_{\rm TRGB}(I)$ changes
by less than $0.1$ mag for $-2.2<[Fe/H]<-0.7$ dex and for ages between
$2$ and $17$ Gyr. For the $K$ band, Ferraro et al.~(\cite{ferr})
derived an empirical relation between $M_{\rm TRGB}(K)$ and the
metallicity in galactic globular clusters.  For metallicities in the
range of the Magellanic Clouds the variation of $M_{\rm TRGB}(K)$ is
about $0.2$ mag; however, this relation might not be valid for
intermediate age populations.  From the theoretical isochrones by
Girardi et al.~(\cite{gir}) the spread of $M_{\rm TRGB}(K)$ is about
$0.3$ mag for ages greater than $2$ Gyr and constant metallicity. This
spread is somewhat less for the $J$ band but it remains higher than
the one derived for the $I$ band. The fact that $M_{\rm TRGB}$ is
modestly sensitive to variations in metallicity and age for the $J$
and $K$ bands implies that the values of $\Delta$ derived in these
bands may not be unbiased estimates of the true difference in distance
modulus between the SMC and the LMC. The $I$ band value should be
better in this respect, but on the other hand, that value is more
sensitive to possible differences in the dust extinction between the
Clouds. So the best estimate of $\Delta$ is probably obtained using
$m_{\rm bol}$, as discussed further in Section~5.

For the  LMC there are several observed  TRGB magnitude determinations
in  the  literature that  can  be compared  to  our  results. Reid  et
al.~(\cite{rmt}) obtained  $m_{\rm TRGB}(I)  = 14.53 \pm  0.05$, after
extinction-correction  with  an   assumed  $A_I=0.07$.  Romaniello  et
al.~(\cite{rscp}) obtained $m_{\rm TRGB}(I)  = 14.50 \pm 0.25$ for the
field  around  SN1987A.  They  corrected each  star  individually  for
extinction,  but found  a mode  of $E(B-V)  = 0.20$  for  their sample
(corresponding  to  $A_I=0.30$).  Sakai et  al.~(\cite{szk})  obtained
$m_{\rm  TRGB}(I) =  14.54 \pm  0.04$. They  also corrected  each star
individually   for  extinction,   but  restricted   their   sample  to
low-extinction  regions with  $A_V  < 0.2$  (corresponding  to $A_I  <
0.10$). The observed value of $m_{\rm TRGB}(I)$ for our sample, $14.54
\pm     0.03$,    is    nicely     consistent    with     all    these
determinations.  However, when  we apply  an extinction  correction of
$A_I=0.27$,   as  appropriate   for   an  assumed   $E(B-V)  =   0.15$
(Section~2.3),  our  corrected  value  falls significantly  below  the
previous determinations. This may  mean that our assumed extinction is
an  overestimate. Support  from  this  comes from  a  recent study  by
Zaritsky  (\cite{zar}). 
He demonstrates that the average extinction towards cool stars
is much lower than for the hotter stars which have typically been used
to estimate the extinction towards the LMC (the latter generally
reside in star-forming regions which are more dusty, among other
things). 
The analysis of  Zaritsky (cf.~his Fig.~12) suggests that  the mode of
the distribution of $A_V$  for stars with temperatures appropriate for
the RGB is as low as  $A_V \approx 0.1$ (corresponding to $A_I \approx
0.05$), but with  a long tail towards higher  extinctions. Either way,
it is  clear that any proper  interpretation of the  TRGB magnitude in
the $I$ band requires an accurate understanding of the effects of dust
extinction. We  have not (yet)  performed such an  extinction analysis
for our  sample, and therefore  refrain from drawing  conclusions from
our $I$ band  results. However, our results are  not inconsistent with
observations  by previous  authors,  provided that  the extinction  is
actually as low as suggested by Zaritsky.

\begin{figure}
\resizebox{7cm}{!}{\includegraphics{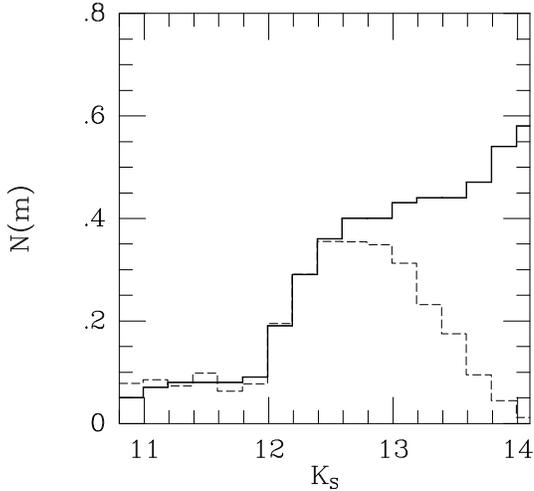}}
\caption{The LMC $K_S$ band magnitude distribution in $0.2$ magnitude bins. 
The dashed curve is for the DENIS data discussed in the present
paper. The solid curve is the histogram obtained from 2MASS data and
presented by Nikolaev \& Weinberg (\cite{Nik}). The abscissa is the
$K_S$ magnitude in the DENIS photometric system. The 2MASS $K_S$
magnitudes were transformed using $K_S($DENIS$) = K_S($2MASS$) - 0.11$,
which was chosen so as to provide the best agreement between the two
histograms. The scale along the ordinate is in arbitrary units.}
\label{NWcomp}
\end{figure}

The best way to circumvent any dependence of the results on
uncertainties in the dust extinction is to go far into the near IR.
There is one very recent determination of $m_{\rm TRGB}$ in the $K_S$
band that can be compared to our results. Nikolaev \& Weinberg
(\cite{Nik}) used data from the 2MASS survey to derive $m_{\rm TRGB}
(K_S) = 12.3 \pm 0.1$ for the LMC, 
without correcting for extinction. For a proper
comparison of this value to our results we must correct for possible
differences in the photometric magnitude systems used by 2MASS and
DENIS. Neither system is identical to the standard CTIO $K$ magnitude
system, but both are quite close. Nikolaev \& Weinberg quote that
their $K_S$ magnitude system agrees with the standard $K$ to within
$0.05$ mag. For the DENIS system the final transformation equations
will not be available until the survey is completed, but the analysis
of Fouqu\'e et al.~(\cite{fou}) yields an absolute flux zero-point (in
Jy) for the DENIS $K_S$ system that differs from the CTIO $K$-band by
$0.08$ mag. Based on this, we do not expect the $K_S$ magnitudes of
2MASS and DENIS to differ by much more than $0.1$ magnitudes. To
determine the actual difference, we compare in Fig.~\ref{NWcomp} our
LMC $K_S$ histogram to that presented by Nikolaev \& Weinberg (using
identical binning). The 2MASS histogram was shifted horizontally to
obtain the best agreement. From this we obtain $K_S($DENIS$) =
K_S($2MASS$) - 0.11 \pm 0.02$. With this photometric correction the
histograms are in good agreement. The slight differences at $K_S<11$ 
magnitudes are probably due to differences in foreground subtraction.
At faint magnitudes the DENIS data become incomplete at brighter
magnitudes than the 2MASS data. However, tests discussed in Appendices
A.3.2 and A.3.3 show that our determinations of $m_{\rm TRGB}$ are not
influenced significantly either by possible incompleteness near the
TRGB or by possible uncertainties in the foreground subtraction. Upon
correction of the Nikolaev \& Weinberg $m_{\rm TRGB}$ determination to
the DENIS $K_S$ magnitude system one obtains $m_{\rm TRGB} (K_S) =
12.19 \pm 0.1$. Somewhat surprisingly, this exceeds our determination
$m_{\rm TRGB}(K_S) = 11.98 \pm 0.04$ by as much as $0.21$
magnitudes. Given that the histograms themselves are in good agreement
(Fig~\ref{NWcomp}), we are forced to conclude that this must be due to
differences in how $m_{\rm TRGB}$ is defined and determined. While we
search for a peak in $N''(m)$ and then add a correction term that is
based on a model, Nikolaev \& Weinberg just determine the peak in the
first derivative $N'(m)$. As discussed in Section~A.1 (see
Fig.~\ref{mag}) this generally yields on overestimate of the actual
TRGB magnitude. Since Nikolaev \& Weinberg do not describe their
analysis technique in detail, it is difficult to estimate the size of
this bias in their result. However, Monte-Carlo simulations that we
discuss in Section~A.4 indicate that it could be $\sim 0.15 \pm 0.06$,
which would explain the observed discrepancy. Note that the same
effect may also affect some of the $I$ band comparisons listed above,
although for those the influence of extinction probably plays the more
significant role.

\section{Distance to the Magellanic Clouds}
To estimate the distance modulus of the Magellanic Clouds we can use
the observed magnitude of the TRGB in either $I$, $J$, $K_S$ or
$m_{\rm bol}$. As discussed in Section~4.4, $I$ has the disadvantage
of being sensitive to uncertain extinction corrections, while $J$ and
$K_S$ have the disadvantage of being sensitive to the assumed
metallicity and age. The most accurate information on the distance is
therefore provided by $m_{bol}$, which is not particularly sensitive to
either dust extinction (cf.~Table~\ref{value}) or metallicity and
age. To quantify the latter we use the stellar evolutionary model
calculations of Salaris \& Cassisi (\cite{saca}). They quantified the
dependence of $M_{\rm TRGB}({\rm bol})$ on the total metallicity
($[M/H]$) of a population, and found that
\begin{equation}
M_{\rm TRGB}({\rm bol}) = -3.949 - 0.178[M/H]+0.008[M/H]^2 ,
\end{equation}
valid for $-2.35<[M/H]<-0.28$ and for ages larger than a few Gyr. 

We determined $[M/H]$ by qualitatively fitting isochrones (Girardi et
al.~\cite{gir}) to the color--magnitude diagram $(K_S, J-K_S)$. We
obtain $Z=0.004\pm 0.002$ for the LMC, in agreement with the value
derived by Nikolaev \& Weinberg, and $Z=0.003\pm 0.001$ for the
SMC. For $Z_\odot = 0.02$ this corresponds to $[M/H]=-0.70$ and
$[M/H]=-0.82$ for the LMC and the SMC, respectively. This in turn
yields $M_{\rm TRGB}({\rm bol}) = -3.82$ for the LMC and $M_{\rm
TRGB}({\rm bol}) = -3.80$ for the SMC. When combined with the results
in Table~\ref{value} we obtain for the LMC that $(m-M) = 18.55 \pm 0.04$
(formal) $\pm 0.08$ (systematic), and for the SMC that $(m-M) = 18.99
\pm 0.03$ (formal) $\pm 0.08$ (systematic). The corresponding
distances are $51$ and $63$ kpc to the LMC and the SMC respectively.

The systematic errors that we quote in our results are the sum in
quadrature of the following possible (identified) sources of error:
(i) $\pm 0.02$ mag due to uncertainties in the phenomenological model
on which the corrections $\Delta m_{2g} (\sigma_{2g})$ are based
(cf.~Section~A.3.1); (ii) $\pm 0.03$ mag to account for the fact that our
assumed average dust extinction of $E(B-V)=0.15$ could plausibly be in
error by $0.05$ (cf.~Table~\ref{value}); (iii) $\pm 0.04$ mag,
reflecting the uncertainties in $M_{\rm TRGB}({\rm bol})$ due to
uncertainties in $[M/H]$; (iv) $\pm 0.04$ mag, reflecting the
uncertainty in $M_{\rm TRGB}({\rm bol})$ at fixed $[M/H]$ suggested by
comparison of the predictions of different stellar evolution models
(Salaris \& Cassisi \cite{saca}; their Fig.~1); 
(v) $\pm 0.05$ mag, being an estimate of the possible systematic error
in our calculation of bolometric magnitudes due to uncertainties in
the underlying spectral model (see Section 2.2).

There have been many previous determinations of the distance modulus
of the LMC, and these have varied widely, from about $18.0$ to $18.7$.
Based on a collection of many determinations, the HST Key Project Team
adopted $(m-M) = 18.50 \pm 0.13$ (Mould et al.~\cite{mould}). Our
determination is in excellent agreement with this value, and actually
has a smaller error. The TRGB method itself has been used previously
by several other authors to study the distance modulus of the LMC, and
our results are consistent with all of these. Reid, Mould \& Thompson
(\cite{rmt}) were the first to apply this technique to the LMC (by
studying the Shapley Constellation III using photographic plates), and
obtained $(m-M) = 18.42 \pm 0.15$. Romaniello et al.~(\cite{rscp})
obtained $(m-M) = 18.69 \pm 0.25$ from a field around SN1987A in the
LMC using HST/WFPC2 data. Sakai et al.~(\cite{szk}) obtained $18.59
\pm 0.09$ from an area of $4 \times 2.7$ square degrees (north of the
LMC bar) studied as part of the Magellanic Cloud Photometric Survey
(Zaritsky, Harris \& Thompson \cite{zht}) using the Las Campanas 1m
telescope.  Nikolaev \& Weinberg (\cite{Nik}) obtained $(m-M) = 18.50
\pm 0.12$ from the subset of 2MASS data that covers the LMC. For the
SMC we are not aware of (recent) TRGB distance modulus measurements,
but our result is consistent with the value $(m-M) = 18.90 \pm 0.10$
quoted by Westerlund (\cite{west}) from a combination of measurements
available in the literature from a variety of techniques.

\section{Conclusions}

We have determined the position of the TRGB for both Magellanic Clouds
using the large statistical sample offered by the DCMC (Cioni et
al.~\cite{cio}). We have presented a new algorithm for the
determination of the TRGB magnitude, which we describe in detail in
the Appendix and test extensively using Monte-Carlo simulations. We
note that any method that searches for a peak in the first derivative
(used by most authors) or the second derivative (used by us) of the
observed luminosity function does not yield an unbiased estimate for
the actual magnitude of the TRGB discontinuity. We stress the
importance of correcting for this bias, which is not generally done.
Our analysis shows that when large enough statistics are available,
contamination by AGB stars does not provide a significant limitation
to the accuracy of the TRGB magnitude determination.

In our analysis we have adopted global values for the extinction of
the Magellanic Clouds and we have derived the metallicity from an
isochrone fit to the giant population to obtain a representative value
for each cloud as a whole. In reality, extinction and metallicity are
likely to vary within each cloud. Clearly, the production of a
detailed extinction map together with precise measurements of the
metallicity is a requirement for a detailed analysis of variations in
structure between different locations within the Clouds, either on the
plane of the sky or along the line of sight. However, such variations
do not influence our distance determinations, which should be accurate
in a globally averaged sense. Uncertainties in the average dust
extinction or metallicity for each cloud are included in the
systematic error budget of our final estimates.

We combine our apparent bolometric TRGB magnitude determinations with
theoretical predictions to derive the distance modulus of the Clouds.
We obtain $(m-M) = 18.55 \pm 0.04$ (formal) $\pm 0.08$ (systematic) for
the Large Magellanic Cloud (LMC), and $(m-M) = 18.99 \pm 0.03$ (formal)
$\pm 0.08$ (systematic) for the Small Magellanic Cloud (SMC). These
results are consistent with many previous studies, including a recent
compilation by Mould et al.~(\cite{mould}). However, only very few
previous studies have yielded determinations of similar accuracy as
those presented here. This re-confirms the TRGB method to be a high
quality method for distance determination of resolved stellar
populations, and stresses the power of large statistical samples in
the NIR such as those provided by the DENIS survey.

\appendix

\section{Determination of the TRGB magnitude: methodology and error analysis}

\subsection{The nature of the TRGB discontinuity}
We  wish  to  determine  the  magnitude $m_{\rm  TRGB}$  of  the  TRGB
discontinuity   from  an   observed  magnitude   distribution  $f_{\rm
obs}(m)$.  In   general,  the   observed  distribution  will   be  the
convolution  of the  intrinsic  magnitude distribution  of the  stars,
$f_{\rm int}(m)$, with some broadening function $E(m)$:
\begin{equation}
   f_{\rm obs}(m) = \int_{-\infty}^{\infty} 
       f_{\rm int}(m') E(m-m') \> {\rm d}m' .
\label{fobs}
\end{equation}
The  function $E(m)$ characterizes  the probability  that a  star with
magnitude $m_0$ is observed to have magnitude $m_{\rm obs} = m_0 + m$.
The shape of  $E(m)$ is generally determined by  the properties of the
observational  errors,  but  other  effects (such  as  differences  in
extinction or distance  among the stars in the  sample) can contribute
as well.

To gain an  understanding of the issues involved  in the determination
of $m_{\rm  TRGB}$ we start by  considering a simple  model. We assume
that $E(m)$ is a Gaussian of dispersion $\sigma$:
\begin{equation}
   E(m) = {1\over{\sqrt{2\pi}\>\sigma}} \> e^{-{{(m/\sigma)^2}\over{2}}} .
\label{gauss}
\end{equation}
We approximate $f_{\rm int}(m)$ by expanding it into a first-order
Taylor expansion near the position of the discontinuity, which yields
\begin{equation}
   f_{\rm int}(m) = \cases{
      f_0 + a_1 (m - m_{\rm TRGB})             , & 
         if $m < m_{\rm TRGB}$ ; \cr
      f_0 + \Delta f + a_2 (m - m_{\rm TRGB})  , & 
         if $m > m_{\rm TRGB}$ . \cr}
\label{taylor}
\end{equation}
The parameters $a_1$ and $a_2$ measure the slope of $f_{\rm int}$ for
magnitudes that are brighter and fainter than $m_{\rm TRGB}$,
respectively. At brighter magnitudes the sample is dominated by AGB
stars, while at fainter magnitudes both AGB and RGB stars contribute.
The parameter $\Delta f$ measures the size of the discontinuity; the
ratio $\Delta f / f_0$ is an estimate of the ratio of the number of
RGB to AGB stars at the magnitude of the RGB tip.

\begin{figure}
\resizebox{7cm}{!}{\includegraphics{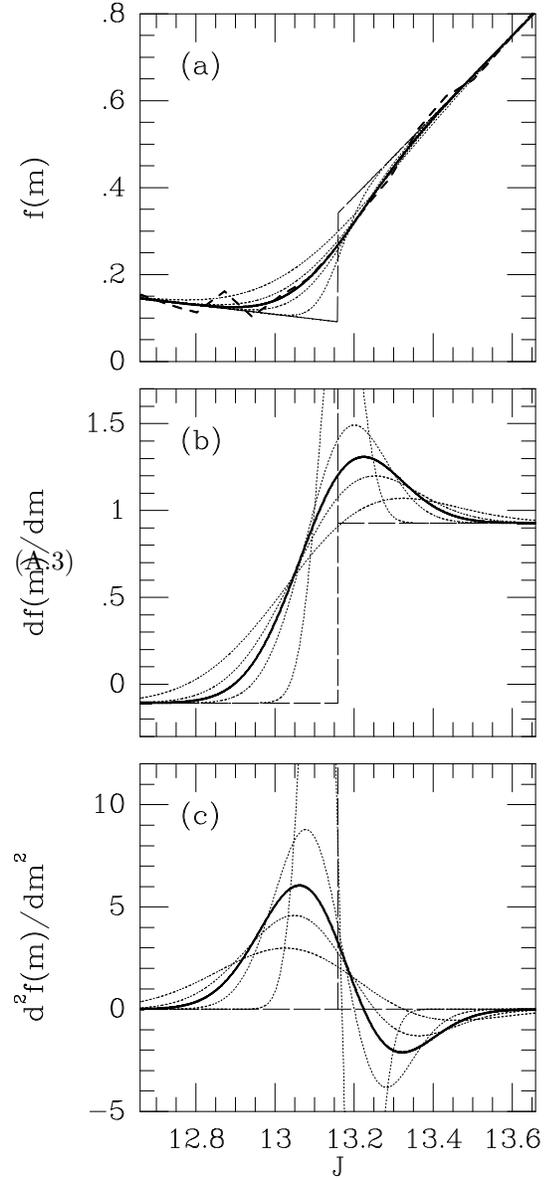}}
\caption{{\bf (a)} The connected heavy dashed curve shows the
foreground-subtracted LMC $J$ band magnitude distribution (thus
providing an expanded view of the region indicated by a rectangle in
the LMC $J$ band panel in Fig.~\ref{tip}) for the expanded sample of
stars detected in the $I$ and $J$ bands (irrespective of whether or
not they were detected in $K_S$). This sample is complete over the
displayed magnitude range.  The heavy solid curve shows the
distribution predicted by the model described in the text. This model
has the intrinsic distribution $f_{\rm int}(m)$ shown as a thin
long-dashed curve, and has an observational convolution kernel $E(m)$
that is a Gaussian with a dispersion $\sigma = 0.126$. For comparison,
thin dotted curves show the predictions obtained when the same
intrinsic distribution $f_{\rm int}(m)$ is convolved with Gaussians of
size $\sigma$ of $0.05$, $0.10$, $0.15$ and $0.20$, respectively. {\bf
(b)} The first derivative of the functions shown in panel (a). {\bf
(c)} The second derivative of the functions shown in panel (a). Note
that the discontinuity at the TRGB induces a peak in both the first
and the second derivative.}
\label{mod}
\end{figure}

We fitted the model defined by Eqs.~(\ref{fobs})--(\ref{taylor}) to
the observed (foreground-subtracted) $J$ band magnitude histogram for
the LMC, which is shown as a connected heavy dashed curve in
Fig.~\ref{mod}a.  The heavy solid curve shows the model distribution
$f_{\rm obs}$ that provides the best fit. The fit is acceptable. The
parameters for this model are: $f_0=0.091$, $\Delta f = 0.250$ (both
in units in which the normalization of $f$ is arbitrary), $a_1 =
-0.108$, $a_2 = 0.928$, $m_{\rm TRGB} = 13.16$ and $\sigma = 0.126$.
The long-dashed curve shows the underlying distribution $f_{\rm
int}(m)$ for this model. For these $J$ band data we know that the
magnitude errors are dominated by photometric zero--point variations
between the scan-strips that constitute the LMC sample (Cioni et
al.~\cite{cio}). These variations have a dispersion of $0.13$ (which
significantly exceeds the formal photometric errors near the TRGB
magnitude, cf.~Fig.~\ref{errs}). In view of this, the value $\sigma =
0.126$ inferred from the model fit is very reasonable.

Model fitting can be used as a general tool to estimate $m_{\rm TRGB}$
from an observed magnitude distribution. However, this technique is
error-prone, since one is essentially solving a deconvolution problem
in which neither the exact shape of the intrinsic magnitude
distribution $f_{\rm int}(m)$ nor that of the kernel $E(m)$ is well
known a priori.  A more robust approach is to locate a feature in the
observed distribution $f_{\rm obs}(m)$ that is a direct consequence of
the discontinuity at $m_{\rm TRGB}$. Since a discontinuity corresponds
(by definition) to an infinitely steep gradient, one obvious approach
is to search for a maximum in the first derivative $f_{\rm obs}'
\equiv {\rm d} f_{\rm obs} / {\rm d} m$.  This approach has been used
in several previous studies of TRGB magnitude determinations (e.g.,
Lee, Freedman \& Madore \cite{lfm}).  For a model with $a_1 = a_2
\equiv a$ one can show that one expects simply $f_{\rm obs}'(m) = a +
\Delta f E(m-m_{\rm TRGB})$, i.e., the first derivative is a Gaussian
centered at $m_{\rm TRGB}$ plus a constant. However, the above
analysis shows that $a_1 \not= a_2$. So while the derivative $f_{\rm
obs}'$ generally does have a maximum near $m_{\rm TRGB}$, the
structure of the first derivative is generally more complicated than a
Gaussian. The heavy curve in Fig.~\ref{mod}b shows $f_{\rm obs}'(m)$
for the model with the parameters determined from the $J$ band data.

The magnitude distribution of stars on the AGB is very different from
that on the RGB.  While the former is approximately constant and in
fact even slightly increasing to brighter magnitudes ($a_1 < 0$), the
latter increases very sharply to fainter magnitudes ($a_2 > 0$).
Hence, not only $f_{\rm int}$, but also its derivative is
discontinuous at $m_{\rm TRGB}$. This corresponds to an infinitely
steep gradient in the first derivative (see the long dashed curves in
Fig.~\ref{mod}), which can be identified by searching for a maximum in
$f_{\rm obs}'' \equiv {\rm d}^2 f_{\rm obs} / {\rm d} m^2$.  For a
model with $\Delta f = 0$ one can show that one expects simply that
$f_{\rm obs}''(m) = (a_2 - a_1) E(m-m_{\rm TRGB})$, i.e., the second
derivative is a Gaussian centered at $m_{\rm TRGB}$.  While the above
discussion shows that the best fit to the data is obtained for $\Delta
f \not= 0$, the value of $\Delta f$ is close enough to zero to ensure
that $f_{\rm obs}''(m)$ is always modestly well approximated by a
Gaussian (especially near its peak).  Fig.~\ref{mod}c shows $f_{\rm
obs}''$ for the model with the parameters determined from the $J$ band
data.

\begin{figure}
\resizebox{9cm}{!}{\includegraphics{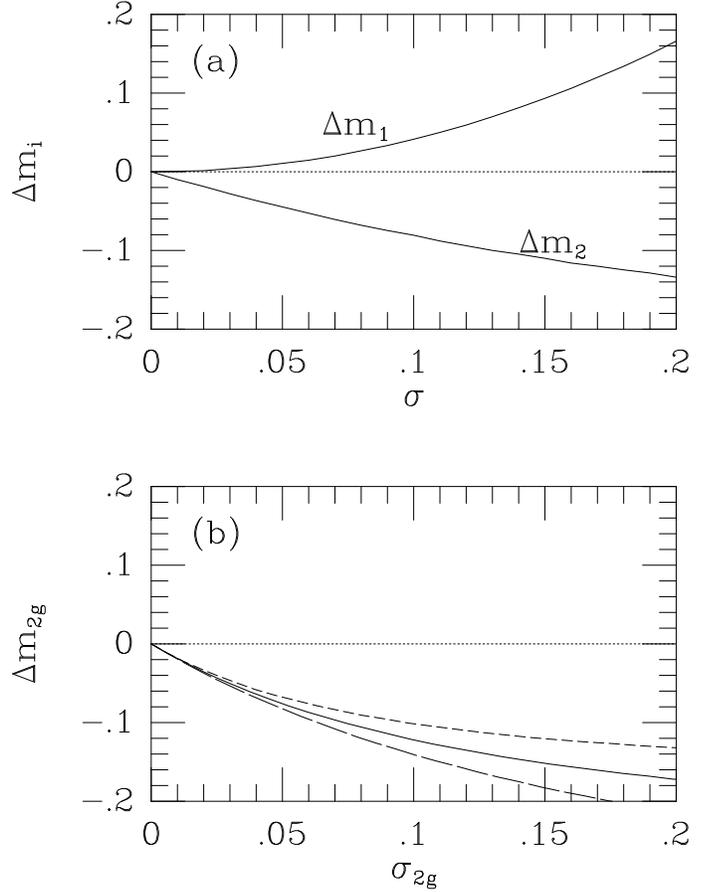}}
\caption{{\bf (a)} The differences $\Delta m_1 \equiv m_1 - m_{\rm
TRGB}$ and $\Delta m_2 \equiv m_2 - m_{\rm TRGB}$ as function of
$\sigma$, for models with the intrinsic magnitude distribution shown
in Fig.~\ref{mod}. The quantities $m_1$ and $m_2$ are, respectively,
the magnitudes at which the first and second derivatives of the
observed magnitude distribution have their peak, while $m_{\rm TRGB}$
is the magnitude of the actual TRGB discontinuity. The quantity
$\sigma$ is the dispersion of the observational convolution kernel
$E(m)$. {\bf (b)} The difference $\Delta m_{2g} \equiv m_{2g} - m_{\rm
TRGB}$ as function of $\sigma_{2g}$, where $m_{2g}$ and $\sigma_{2g}$
are the mean and dispersion of the Gaussian that best fits the peak in
$f_{\rm obs}''(m)$.  The solid curve refers to the same models as in
(a), and provides the correction term that we have applied to the
observed $m_{2g}$ to obtain estimates of $m_{\rm TRGB}$. The other
curves are for models with $\Delta f = 0.18$ (dashed) and $\Delta f =
0.38$ (long-dashed) in Eq.~\ref{taylor}; as discussed in
Section~A.3.1, the differences between these curves and the solid
curve provide an estimate of possible systematic errors in our results
due to uncertainties in the adopted model for $f_{\rm int}(m)$.}
\label{mag}
\end{figure}

While the discontinuity in $f_{\rm int}$ causes both a maximum in
$f_{\rm obs}'$ at a position $m_1$ and a maximum in $f_{\rm obs}''$ at
a position $m_2$, it is important to realize that neither provides a
unbiased estimate of $m_{\rm TRGB}$. Fig.~\ref{mag}a shows for the
model derived from the $J$ band data the differences $\Delta m_1
\equiv m_1 - m_{\rm TRGB}$ and $\Delta m_2 \equiv m_2 - m_{\rm TRGB}$
as function of $\sigma$.  In absolute value, the differences increase
monotonically with $\sigma$. The value of $m_1$ always provides an
overestimate of $m_{\rm TRGB}$ while $m_2$ always provides an
underestimate. It is important to realize that in practice, because of
finite statistics, one must always apply a certain amount of smoothing
to real data to obtain an adequate estimate of either $f_{\rm obs}'$
or $f_{\rm obs}''$. This smoothing usually takes the form of binning
(e.g., Lee, Freedman \& Madore \cite{lfm})) or kernel smoothing (e.g.,
Sakai, Madore \& Freedman \cite{sakai}). When assessing the size of
the bias terms in Fig.~\ref{mag}a for any particular application, the
value of $\sigma$ along the abscissa should therefore not be taken
merely as the average photometric error for the data, but should
include the effect of the additional smoothing that was applied to
obtain the estimate of either $m_1$ or $m_2$. While photometric errors
of a few hundredths of a magnitude are often routinely achieved, the
additional smoothing or binning applied during data processing is
often as large as 0.1 to 0.2 magnitudes. According to Fig.~\ref{mag}a,
this can induce systematic biases in the estimate of $m_{\rm TRGB}$
that are of the same order. So while this is not typically done (e.g.,
Sakai, Zaritsky \& Kennicutt \cite{szk}; Nikolaev \& Weinberg
\cite{Nik}), we do believe that such systematic biases should
be calculated and corrected for.

Previous authors have generally searched for the magnitude of the TRGB
by determining the position of the peak in $f_{\rm obs}'$. As far as
we know, no one has yet used $f_{\rm obs}''$. This is presumably for
the obvious reason that it is more difficult to determine the second
derivative from noisy data than the first derivative.  However, the
situation for the DCMC catalogue differs considerably from that for
most other studies. First, we have a very large number of stars, so
that it is actually not a problem to accurately determine $f_{\rm
obs}''$. Second, the random errors in the sample are relatively
large. This is not because of photometric errors (which are small,
cf.~Fig.~\ref{errs}) but because of photometric zero--point variations
between the scan-strips that constitute the sample. The effect of the
size of the errors on the properties of $f_{\rm obs}'$ and $f_{\rm
obs}''$ are illustrated by the dotted curves in Fig.~\ref{mod}, which
show predictions for the same model as before, but for values of
$\sigma$ of $0.05$, $0.10$, $0.15$ and $0.20$, respectively. We have
found that the values of $\sigma$ appropriate for our analysis are
such that the peak in $f_{\rm obs}'(m)$ is generally not the most
easily recognizable feature in the data.  After extensive testing we
concluded that for our data $f_{\rm obs}''(m)$ provides a better
handle on $m_{\rm TRGB}$ than does $f_{\rm obs}'(m)$.

In practice, we estimate the properties of the peak in $f_{\rm
obs}''(m)$ by performing a Gaussian fit. This yields $m_{2g}$, the
center of the best-fitting Gaussian, and $\sigma_{2g}$, the dispersion
of the best-fitting Gaussian (in general, the value of $\sigma_{2g}$
is roughly of the same order as $\sigma$, and $\Delta m_{2g}$ is
roughly of the same order as $\Delta m_2$). For given $f_{\rm int}$,
both $m_{2g}$ and $\sigma_{2g}$ are unique monotonic functions of
$\sigma$. So one can view $\Delta m_{2g} \equiv m_{2g} - m_{\rm TRGB}$
to be a function of $\sigma_{2g}$. The solid curve in Fig.~\ref{mag}b
shows this function for the $f_{\rm int}$ parameterization derived
from the $J$ band data.
  
\subsection{Implementation and formal errors} 
To implement our  strategy we bin the observed  stellar magnitudes for
the region of the sky of  interest into a histogram, using a fixed bin
size $b$. As described in Section~3.1, we do the same for observations
of an  offset field, and  subtract an appropriately scaled  version of
the offset field  histogram from the main field  histogram to obtain a
foreground-subtracted histogram $N(m)$. We then apply a Savitzky-Golay
filter  (e.g.,  Press et  al.~\cite{press})  to  estimate the  second
derivative  ${\rm d}^2  N(m) /  {\rm d}m^2$  at the  position  of each
bin. This yields for bin number $i$
\begin{equation}
  [{\rm d}^2 N / {\rm d}m^2]_i = \sum_{j=-J}^{J} c_j \> [N(m)]_{i+j} ,
\end{equation}
where the  $c_j$ are Savitzky-Golay coefficients for  the chosen value
of  $J$ and  the desired  derivative order  $L=2$. The  filter  fits a
polynomial of order  $M$ to the data points $[N(m)]_j$  with $j = i-J,
\ldots, i+J$,  and then evaluates  the $L^{\rm th}$ derivative  of the
polynomial at bin $i$ to estimate $[{\rm d}^2 N / {\rm d}m^2]_i$. Once
a histogram  approximation to  $[{\rm d}^2 N  / {\rm d}m^2]$  has been
calculated, we  search for  a peak  and fit a  Gaussian in  the region
around the  peak to  obtain $m_{2g}$ and  $\sigma_{2g}$ (the  mean and
dispersion  of  the  best-fitting  Gaussian).  From  these  values  we
estimate the magnitude $m_{\rm TRGB}$ as
\begin{equation}
  m_{\rm TRGB} = m_{2g} - \Delta m_{2g} (\sigma_{2g}) ,
\label{corfac}
\end{equation}
where the correction term  $\Delta m_{2g} (\sigma_{2g})$ is taken from
Fig.~\ref{mag}b. To  summarize, $m_{\rm  TRGB}$ is  estimated  as the
position where the second derivative of the observed histogram has its
maximum, plus  a small  correction that  is based on  a model  for the
underlying magnitude distribution $f_{\rm int}$.

We performed extensive Monte-Carlo simulations to assess the accuracy
of the $m_{\rm TRGB}$ estimates produced by this algorithm. In these
simulations Cloud stars are drawn from the magnitude distribution
$f_{\rm int}$ given by Eq.~(\ref{taylor}), using as before the
parameters determined from the $J$ band data. Foreground stars are
drawn from a smooth magnitude distribution that matches that inferred
from our data, both for the main field and a hypothetical offset
field.  To each stellar magnitude an error is added that is drawn from
a Gaussian with dispersion $\sigma$.  The numbers of stars in the
simulations were chosen to match those in our datasets. In each
simulation, the magnitudes thus generated are analyzed in exactly the
same way as the real data to obtain $m_{2g}$ and $\sigma_{2g}$, and
from these (using Eq.~\ref{corfac}) an estimate ${\tilde m}_{\rm
TRGB}$.  This procedure is then repeated many times in Monte-Carlo
fashion, and for the resulting ensemble we calculated the mean
$\langle {\tilde m}_{\rm TRGB} \rangle$ and dispersion $\sigma_{m,{\rm
TRGB}}$ of the ${\tilde m}_{\rm TRGB}$ estimates, as well as the mean
$\langle \sigma_{2g} \rangle$ of the $\sigma_{2g}$. In the simulations
we experimented with the choice of the algorithm parameters $b$, $J$,
and $M$. 
We found that accurate results were obtained with, e.g., $J=3$, $M=2$
and a binsize $b = 0.07$ magnitudes. These parameters were therefore
generally adopted for the further analysis (with the exception of the
SMC $K_S$ band data, for which we used the slightly larger bin size $b
= 0.10$ magnitudes). The Savitzky-Golay coefficients for this choice
of parameters are $c_j = {\bar c}_j / b^2$, with ${\bar c}_{0} =
-0.0476$, ${\bar c}_{1} = {\bar c}_{-1} = -0.0357$, ${\bar c}_{2} =
{\bar c}_{-2} = 0$, ${\bar c}_{3} = {\bar c}_{-3} = 0.0595$. With
these parameters we found that $|\langle {\tilde m}_{\rm TRGB} \rangle
- m_{\rm TRGB} | < 0.01$ magnitudes, independent of the assumed
$\sigma$. Hence, the algorithm produces unbiased estimates of $m_{\rm
TRGB}$. This result was found to be rather insensitive to the precise
choice of the algorithm parameters; different parameters generally
yielded similar results for $m_{\rm TRGB}$.
The formal error on a determination of $m_{\rm TRGB}$ from
real data is obtained as follows: (i) we run simulations with the
appropriate numbers of stars, for a range of $\sigma$ values; (ii) we
identify the value of $\sigma$ that yields a value of $\langle
\sigma_{2g} \rangle$ that equals the value of $\sigma_{2g}$ inferred
from the data; (iii) the corresponding value of $\sigma_{m,{\rm
TRGB}}$ is the formal error that was sought. The errors thus inferred
are listed in Table \ref{value}; typical values are $0.02$--$0.05$
magnitudes.

\subsection{Assessment of systematic errors}
The Monte-Carlo simulations provide accurate estimates of the formal
errors in the $m_{\rm TRGB}$ determinations due to the combined
effects of the finite number of stars and the properties of our
adopted algorithm. However, they provide no insight into possible
systematic errors.  We have performed a number of additional tests to
assess the influence of possible sources of systematic errors.

\subsubsection{Accuracy of the correction term $\Delta m_{2g}$} 
Our estimates for $m_{\rm TRGB}$ are obtained from Eq.~(\ref{corfac}),
in  which we add  to the  observed magnitude  $m_{2g}$ of  the $f_{\rm
obs}''(m)$ peak  a correction $\Delta  m_{2g}$ that is derived  from a
model.  Any error  in the  model  will change  the correction  $\Delta
m_{2g}$,  which in  turn  yields  a systematic  error  in the  derived
$m_{\rm TRGB}$.  It is therefore important to  understand the accuracy
of the model.

There  are  two  main  parameters  in fitting  the  model  defined  by
Eqs.~(\ref{fobs})--(\ref{taylor}) to  an observed histogram, namely the
`step-size'  $\Delta f$  of  the function  $f_{\rm  int}(m)$, and  the
dispersion $\sigma$ of the convolution kernel $E(m)$. These parameters
are highly correlated. If (as  compared to the best fit model) $\Delta
f$ is increased, then an appropriate simultaneous increase in $\sigma$
will yield a predicted profile  $f_{\rm obs}(m)$ that is only slightly
altered. From experiments with our Monte-Carlo simulations we conclude
that for  all $0.18 \leq \Delta f  \leq 0.38$ one can  still obtain an
acceptable fit  to the observed  $J$ band magnitude histogram.  At the
lower end  of this range we require  $\sigma = 0.105$ and  at the high
end $\sigma = 0.169$, neither  of which seems entirely implausible for
the  $J$ band  data. The  dashed curves  in Fig.~\ref{mag}b  show the
correction   factors   $\Delta   m_{2g}   (\sigma_{2g})$   for   these
models. These  can be compared to  the solid curve,  which pertains to
the model  with $\Delta f =  0.25$ shown in Fig.~\ref{mod}. A typical
value of  $\sigma_{2g}$ for our  data is $\sim 0.11$.  Fig.~\ref{mag}b
shows  that for  this $\sigma_{2g}$  the systematic  error  in $\Delta
m_{2g}$ (and hence $m_{\rm TRGB}$)  due to uncertainties in $\Delta f$
is approximately $0.02$ magnitudes.

The correction term $\Delta m_{2g} (\sigma_{2g})$ that we have applied
to all our data was derived from LMC data in the $J$ band. This would
not be adequate if the shape of $f_{\rm int}(m)$ differs significantly
among the $I$, $J$ and $K_S$ bands, or among the LMC and the
SMC. However, visual inspection of Fig.~\ref{tip} does not strongly
suggest that this is the case: the shape of the observed magnitude
histograms near the TRGB is similar in all cases.  Quantitative
analysis supports this, and demonstrated that values of $0.18 \leq
\Delta f \leq 0.38$ are adequate for all our data.

\subsubsection{Incompleteness}
In our main sample we have only included stars that were confidently
detected in all three photometric bands. Fig.~\ref{tip} shows that for
this sample incompleteness starts to be an issue at brightnesses that
are only a few tens of a magnitude fainter than the inferred $m_{\rm
TRGB}$.  One may wonder whether this could have had a systematic
influence on the $m_{\rm TRGB}$ determinations. To assess this we
applied our algorithm also to a different (extended) sample consisting
of those stars that were detected in the $I$ and $J$ bands
(irrespective of whether or not they were detected in $K_S$), which is
complete to much fainter magnitudes than the main sample (heavy dashed
curves in Fig.~\ref{tip}). The RMS difference between the $m_{\rm
TRGB}$ estimates from the main and the extended sample (for those
cases where both are available) was found to be $0.04$, which can be
attributed entirely to the formal errors in these estimates. We
therefore conclude that there is no evidence for systematic errors due
to possible incompleteness.

\subsubsection{Foreground subtraction}
Our method for foreground subtraction (see Section~3.1) is based on an
empirical scaling of the magnitude histogram for an offset field.  To
assess the effect of possible uncertainties in the foreground
subtraction we have, as a test, done our analysis also without any
foreground subtraction (i.e., using the thin solid curves in the
$N(m)$ panels of Fig.~\ref{tip}).  Even this very extreme assumption
was found to change the inferred $m_{\rm TRGB}$ values only at the
level of $\sim 0.02$, which can be attributed entirely to the formal
errors in the estimates. We therefore conclude that there is no
evidence for systematic errors due to uncertainties in the foreground
subtraction.

\subsubsection{Extinction}
Extinction enters into our analysis in various ways. For the I, J and
$K_S$ data we have performed our analysis on data that were not
corrected for extinction.  Instead, we apply an average extinction
correction to the inferred $m_{\rm TRGB}$ values after the analysis.
Obviously, any error in the assumed average extinction for the sample
translates directly into an error in $m_{\rm TRGB}$. Table \ref{value}
lists for each band the shift in $m_{\rm TRGB}$ that would be
introduced by a shift of $+0.05$ in the assumed $E(B-V)$ (a shift of
$-0.05$ in the assumed $E(B-V)$ would produce the opposite shift in
$m_{\rm TRGB}$). It should be noted that our analysis does not assume
that the extinction is constant over the region of sky under study.
If there are variations in extinction then this causes an additional
broadening of the convolution kernel $E(m)$ beyond what is predicted
by observational errors alone. The width of the convolution kernel is
not assumed to be known in our analysis, but is calibrated indirectly
through our determination of $\sigma_{2g}$ (the dispersion of the
$f_{\rm obs}''(m)$ peak). Hence, any arbitrary amount of extinction
variations within the Clouds will neither invalidate our results, nor
increase the formal errors.

In our calculation of the bolometric magnitudes $m_{\rm bol}$ of the
individual stars in our sample from the observed $J$ and $K_S$
magnitudes we do correct for extinction. The effect of a change in the
assumed $E(B-V)$ affects the inferred $m_{\rm TRGB}$ values in a
complicated way, because both the magnitudes and the colors of
individual stars are affected.  We therefore performed our entire
analysis of the $m_{\rm bol}$ histograms for three separate assumed
values of $E(B-V)$, namely $0.10$, $0.15$ and $0.20$.  From these
analyses we conclude that an increase in $E(B-V)$ of $+0.05$ decreases
the inferred bolometric $m_{\rm TRGB}$ by $-0.03$ (a shift of $-0.05$
in the assumed $E(B-V)$ would produce the opposite shift in $m_{\rm
TRGB}$). As for the $I$, $J$ and $K_S$ data, extinction variations
within the Clouds will not invalidate the results or increase the
formal errors.

\subsection{Comparison to other methods}

Most  authors have  searched for the  magnitude $m_1$  of the
peak in the first derivative  $f_{\rm obs}'$ to estimate the magnitude
$m_{\rm TRGB}$ of  the TRGB discontinuity.  While this  is a perfectly
good approach, it is important to realize that this by itself does not
yield an  unbiased estimate  of $m_{\rm TRGB}$.  This was  pointed out
previously   by   Madore   \&   Freedman   (\cite{mado};   see   their
Fig.~3).  However, they  were not  overly concerned  with  this, since
their aim was to test the limitations on determining $m_{\rm TRGB}$ to
better  than $\pm  0.2$  mag. As  a  result, it  has  not been  common
practice  to estimate  the bias  $\Delta m_1$  intrinsic to  $m_1$ and
correct for  it. Fig.~\ref{mag}a also  shows that for small  values of
$\sigma$ one has $|\Delta m_1|  < |\Delta m_2|$, so the application of
a correction  may seem less important  for methods based  on the first
derivative  than for  those based  on the  second derivative.   On the
other hand, it has now  become possible to determine $m_1$ with formal
errors of order $0.1$ mag  or less (e.g., Sakai, Zaritsky \& Kennicutt
\cite{szk}; Nikolaev  \& Weinberg \cite{Nik}),  so it is  important to
correct for systematic  biases even if one uses  the first derivative,
as we will illustrate.

To estimate quantitatively the size of possible biases in the
results of  previous authors one  must do Monte-Carlo  simulations for
their  exact  observational setup  and  analysis  procedure, which  is
beyond the scope of the  present paper. However, as an illustration it
is useful to consider the result of Nikolaev \& Weinberg (\cite{Nik}),
who find  from 2MASS data for  the LMC that $m_{\rm  TRGB}(K_S) = 12.3
\pm 0.1$.  This corresponds to  $m_{\rm TRGB}(K_S) = 12.19 \pm 0.1$ in
the DENIS  photometric system, which conflicts  significantly with our
result    $m_{\rm    TRGB}(K_S)     =    11.98    \pm    0.04$    (see
Section~4.4).  Nikolaev  \&  Weinberg  derived their  result  from  an
analysis of the derivative of the observed magnitude distribution; the
latter  is shown and  listed as  a histogram  with $0.2$  mag.~bins in
their Fig.~9 and Table~1. If they used the Sobel edge detection filter
suggested by Madore \&  Freedman (\cite{mado}) on this histogram, then
Monte-Carlo  simulations  that  we  have  done (similar  to  those  in
Section~A.2) indicate that their  estimate of $m_1$ could overestimate
$m_{\rm TRGB}$ by as much as $\sim 0.15 \pm 0.06$. If we correct their
result for  this bias, then we  obtain $m_{\rm TRGB}(K_S)  = 12.04 \pm
0.12$ for their data, in good agreement with our result. Romaniello et
al.~(\cite{rscp})  use a  bin size  as large  as $0.25$  mag  in their
analysis, and their estimate of the TRGB magnitude is therefore likely
to be biased upward even more.

Our method differs from that employed by Sakai, Madore \& Freedman
(\cite{sakai}) in that they employ kernel smoothing and estimate
$f_{\rm obs}'$ as a continuous function, while we employ
histograms. Sakai et al.~quote as an advantage of their technique that
it avoids the arbitrary choice of bin size and histogram starting
point. While this is true, we have not found any evidence that this
makes a significant quantitative difference. Our Monte-Carlo
simulations indicate that our results obtained from histograms are
unbiased to better than 0.01 mag., and we have found this to be true
for all histogram starting points and a large range of reasonable bin
sizes. However, we should point out that for this to be the case it is
important to apply appropriate corrections for systematic biases
(which applies equally to histograms estimates and kernel smoothing
estimates).

A final issue worth mentioning is the estimation of the formal error
in $m_{\rm TRGB}$. We have done this through Monte-Carlo simulations,
which is probably the most robust way to do this. By contrast, Sakai,
Zaritsky \& Kennicutt (\cite{szk}) quote as the formal error the FWHM
of the observed peak in $f_{\rm obs}'$. It should be noted that this
is not actually accurate (it is probably conservative). Recall from
Section~A.1 that for the simplified case in which $a_1 = a_2 \equiv a$
in Eq.~(\ref{taylor}), one has $f_{\rm obs}'(m) = a + \Delta f
E(m-m_{\rm TRGB})$. Hence, the dispersion of the peak in $f_{\rm
obs}'(m)$ measures the random error in the individual stellar
magnitude measurements (plus whatever smoothing was applied to the
data). This dispersion is independent of the number of stars in the
sample ($N$), and therefore cannot be a measure of the formal error in
$m_{\rm TRGB}$. The true formal error (i.e., the dispersion among the
results obtained from different randomly drawn samples) scales with
the number of stars as $1/\sqrt{N}$.

\end{document}